\documentclass[a4paper,floatfix,rmp,twocolumn,showkeys,superscriptaddress]{revtex4}
\usepackage{graphicx}
\usepackage{epstopdf}
\begin{document}


\title{Emergence of multicellularity in a model of cell growth, death and aggregation under size-dependent selection}

\providecommand{\ICREA}{ICREA-Complex Systems  Lab, Universitat Pompeu
  Fabra,   Dr    Aiguader   88,   08003   Barcelona,   Spain}
\providecommand{\SFI}{Santa Fe  Institute, 1399 Hyde  Park Road, Santa
  Fe NM 87501, USA}
\providecommand{\IBE}{Institut de Biologia Evolutiva, UPF-CSIC, Psg Barceloneta 37, 08003 Barcelona, Spain}

\author{Salva Duran-Nebreda}     
\affiliation{\ICREA}   
\affiliation{\IBE} 

 \author{Ricard Sol\'e\footnote{corresponding   author}}   \affiliation{\ICREA}
\affiliation{\IBE}
\affiliation{\SFI}
 
\vspace{0.4 cm}
\begin{abstract}
\vspace{0.2 cm}
How multicellular life forms evolved out from unicellular ones constitutes a major problem 
in our understanding of the evolution of our biosphere. A recent set of experiments involving 
yeast cell populations has shown that selection for faster sedimenting cells leads to the appearance of 
stable aggregates of cells that are able to split into smaller clusters. 
It was suggested that the observed evolutionary patterns could be the result of 
evolved programs affecting cell death. 
Here we show, using a simple model of cell-cell interactions and evolving adhesion rates, 
that the observed patterns in cluster size and localized mortality can be easily interpreted in terms of 
waste accumulation and toxicity driven apoptosis. 
This simple mechanism would have played a key role in the early evolution of  multicellular life forms based on both
aggregative and clonal development. The potential extensions 
 of this work and its implications for natural and synthetic multicellularity are discussed. 
\end{abstract}


\maketitle


\section{Introduction}


One of the key major transitions of evolution involved the emergence of multicellular life forms 
out from single-cell systems [1,2]. The standard view 
is that groups of cooperating cells are able to take advantage of division of labor in order to better exploit 
external resources, avoid predators  or improve given adaptive traits [3,4]. 
Yet the transition multicellularity (MC) encapsulated in this picture involves an increase 
in overall complexity [5] 
and thus increasing costs for coordinated cooperating behavior. The main problem is then to understand 
what makes the tradeoff between these two sides balance out.

Available phylogenetic techniques have shed light into how and when the roots of multicellularity got 
established [6-9]. Particularly, comparative analyses of different clades of multicellular organisms 
have proven to be very useful in delineating of the genetic toolkit required for multicellular existence [10]. 
These studies show that cell-cell communication and adhesion genes were co-opted from ancestral functions
unrelated to multicellular phenotypes into robust developmental processes. In this vein, many unicellular species 
have the potential to behave (at least in some circumstances) as cooperative ensembles of 
cells [11,12]. 

Two major paths towards MC have been identified [7]. 
The first is clonal development [6,8] which involves the evolution of 
a life cycle that requires all cells to display adhesion molecules capable of 
maintaining them together and for all cells to share the same genotype. The second is aggregative development. 
This alternative path does not require clonality and is present is some well known but rare systems, 
such as slime molds [1]. In this scenario, MC aggregates 
can form under some conditions and disaggregate into non-clonal individual cells. More recently, 
it has been found that some unicellular species display a MC pattern of development based 
on aggregative dynamics [13]. These remarkable findings suggest 
that non-clonal developmental processes might have played an important role in the early evolution 
of multicellular life forms [14].

\begin{figure*}
\includegraphics[width=0.75 \textwidth]{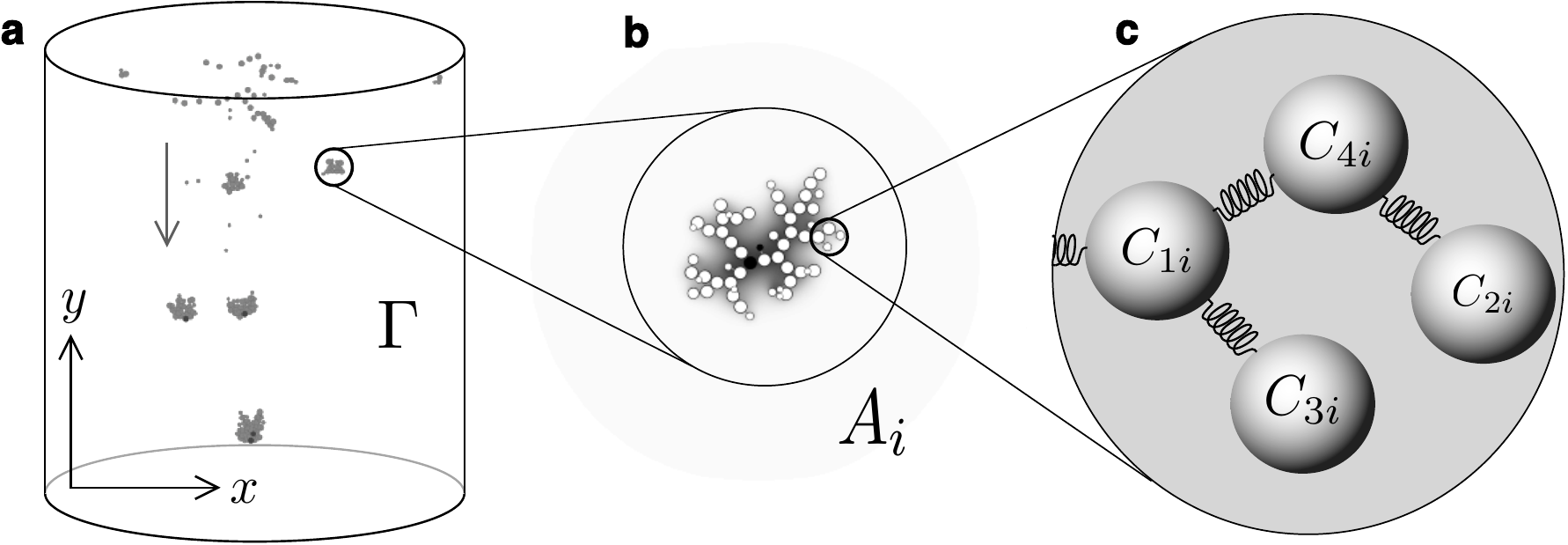}
\caption{Modeling evolution of multicellular aggregates. Following the experimental 
setup described in (Ratcliff et al 2012) we consider a physically embodied description 
of aggregates growing and falling under the action of gravity (a). For simplicity, the 
spatial domain is confined to a two-dimensional lattice. In it,
yeast cells have a limited number of potential attached cells and, in response to the local 
concentration of chemical species, cells can divide or die.
They can remain attached to daughter cells due to failure of separation, thus forming aggregates (b). 
Such aggregates are modeled in terms of simple repelling particles connected by springs (c). 
The physical displacement or breakage of these aggregates is introduced 
by cell death (see text).}
\end{figure*}

In a recent set of experiments [15-18] artificial selection of cell clusters under gravity constraints
was performed. The authors took advantage of the fastest sedimentation speed of cell aggregates 
of increasing size as a shortcut for selecting for more complex cell assemblies and potential 
mutations favouring them. Remarkably, after a relatively short number of generations, obtained by 
repeated culture transfers, the so called snowflake phenotypes appeared in a predictable way. 
These are rounded clusters of cells that appear attached to each other. The authors also studied 
the role played by cellular interactions and cluster structure on the underlying 
reproductive processes. It was found that clusters do not reproduce through events associated to 
single cells but instead involved a cluster-level set of events and -it was argued- a division of labor 
resulting from an apparently active control of apoptosis. The sequence of events as reported from 
this microcosm experiments has important consequences for our understanding of the 
evolution of multicellularity and potential 
scenarios for recreating the first steps from single cells to cooperating ensembles 
and organisms. The claim that evolved apoptotic paths might be at work is specially appealing.

Performing actual experiments involving physical aggregates is a necessary step towards 
reconstructing the events that pervaded the rise of MC. Most theoretical models consider 
genetic traits but typically ignore embodiment: both individual cells and aggregates are mapped into 
non-dimensional, point objects, but including the actual embodiment makes a difference [19]. 
In this paper we present a computational model of Ratcliff's et al experiments, by 
dealing with a simple set of assumptions that support an alternative interpretation, based on the
 accumulation of toxic products -such as acetic acid or ammonia- inherent to yeast metabolism [20,21], 
 which could take place inside a large 
 cluster instead of programmed cell death [22]. The model involves 
a physical, embodied implementation of cellular aggregates falling in a given medium. Our model 
allows to reproduce the basic experimental results and provide a computational framework to 
analyse alternative scenarios for the emergence of MC.

\section{Methods}

The experiments summarised above include a selection process obtained by sequentially growing yeast
in a well mixed medium and selecting for the cells displaying faster sedimentation. This approach immediately 
makes larger clusters of cells to be preferentially selected as in [15]. 
Here we examine these results under the light of a simple, embodied computational model using 
the NetLogo programming language, which allows 
to simulate Newtonian physics [23] on groups of interacting particles. 
Here cells are represented as objects having a given position and velocity. Cell-cell interactions are 
modelled by simple but physically meaningful spring-like interactions. Similarly, the interaction between 
cells and the fluid environment within which they move (essentially under free fall) is also introduced 
in a realistic manner. Additional rules related to nutrient and waste diffusion and consumption are also introduced.

\subsection{Computational model}

Our model considers a spatially extended description of the individuals and their interactions (figure 1a-c). 
For simplicity, we assume a two-dimensional spatial domain $\Gamma$. In this area, cells are described
 as point physical objects interacting (figure 1b), when attached to each other, through springs (figure 1c).  
Moreover, these objects are subject to gravitation fields when appropriate, or display a random walk otherwise 
(see selection process section).

The experiment starts with a population of single cells located on random positions along $\Gamma$. 
Cells increase in biomass through the consumption of the available nutrient to them and, if a particular
threshold is surpassed, a cell can divide and asymmetrically split the resources between the two resulting cells.
Stochastically, these two new cells can fail to separate correctly and become an aggregate, 
which in turn determines some of the individual properties of the cells (namely the sedimentation speed). 
Yeast cells are considered to have a limited number of potential attached cells due to geometrical constraints. 
As such, aggregates in the simulation are, in essence, Bethe lattices with $k_n$ neighbors (we consider 
$k_n=4$ as the upper limit due to physical constraints). 

Following the original set up [15], the simulated experiments include two distinct phases: growth and sedimentation. 
In the former, cells are grown in a well mixed tank until a certain number is reached. In this phase, cells 
move by random walking through $\Gamma$, consume nutrients in order to grow and multiply, 
but also generate generic waste byproducts that can cause their death. In the second stage, cells fall 
under the action of a gravitational field, modelled by a biased random walk using Stoke's law for the 
vertical component of the bias. This selection step is considered sufficiently short so that cells neither divide 
nor die. After a given time -the settling time-, those aggregates collected at the lower part of $\Gamma$ are 
used to seed back the next round of the process, to be located again randomly all over the spatial domain. 

The basic components of the models presented here are cells or clusters of cells resulting 
from birth and death processes. At any time $t$ in a given simulation, the total population 
will be composed by a set $\cal A$ of 
$n(t)$ aggregates, namely 
\begin{equation}
{\cal A} = \{ A_1, ..., A_{n(t)} \}.
\end{equation}
Each aggregate $A_i$ is formed by a set of linked cells, i. e. 
\begin{equation}
A_i = \{ C_{1i}, C_{2i}, ..., C_{n_i,i} \}.
\end{equation}
Let us label as $|A_i|$ the size of the i-th aggregate. The mass of each ($i$) cell within 
a given ($j$) aggregate will be indicated as $M_{ij}$. Cells in the model have a constant 
uptake of resources from their immediate sorroundings. At the same time nutrient diffuses
and is homogeneously replenished in $\Gamma$. To take into account these processes, 
nutrient concentration change in the finite                 element $\phi_{ij}$  is given by the following partial differential equation:

\begin{figure}[htbp]
\begin{center}
\includegraphics[width=0.45 \textwidth]{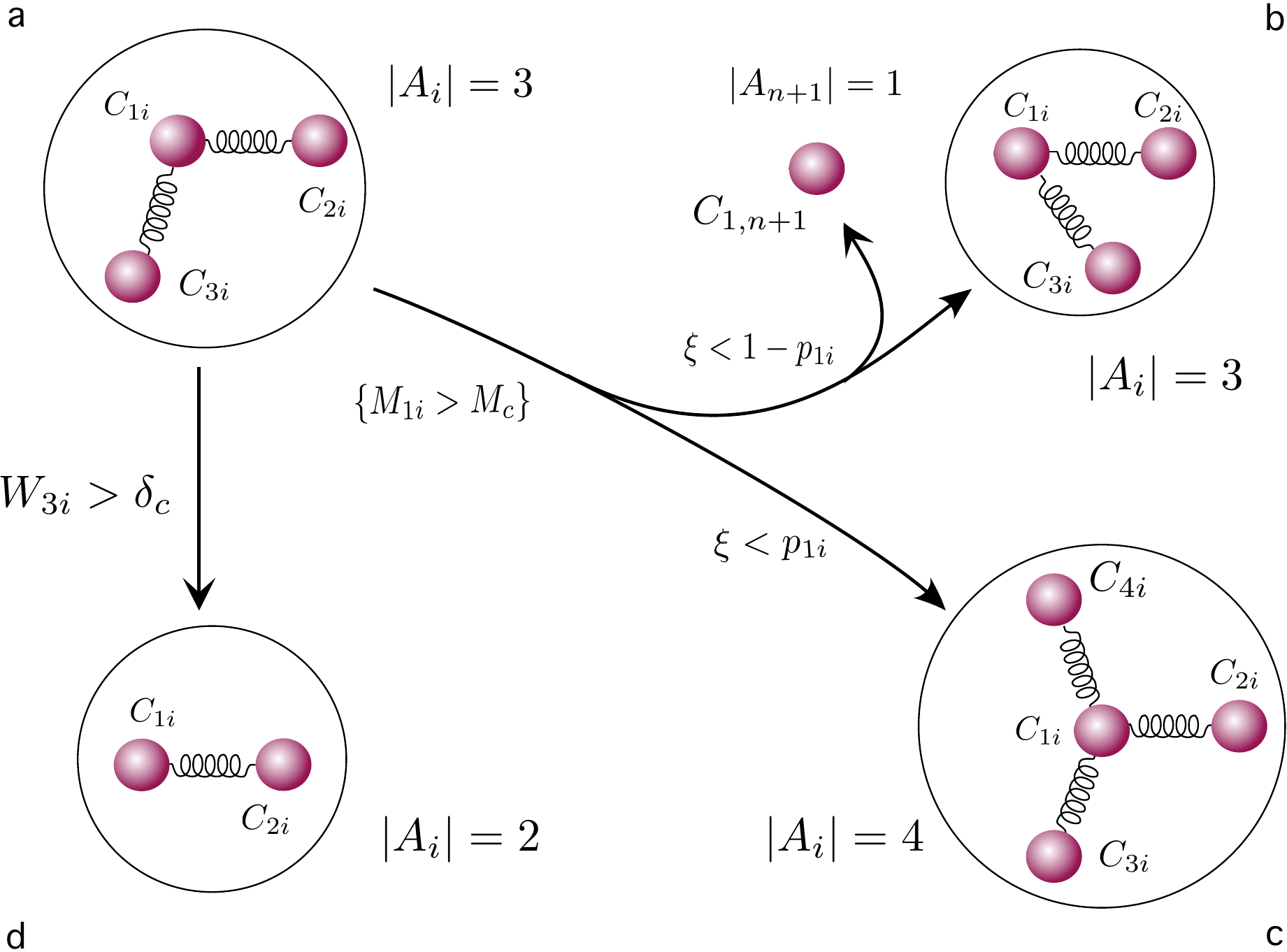}
\caption{The basic set of rules used in our model approach to the evolution experiments. The model
introduces a cellular death mechanism based on metabolic byproduct accumulation. A given aggregate $A_i$, here 
composed by just three cells, is shown in (a). It can experience three different types of processes: cell 
division without (b) and with (c) an increase of aggregate size and (d) cell death. The last scenario 
takes place if the waste concentration of -say- the third cell $C_{3i}$ is above a critical threshold $\delta_c$. 
If the first cell, $C_{1i}$, has a mass larger than another threshold $M_c$ and has fewer than 4 
spring-connected relatives, it will split generating an additional cell. 
This new cell can leave the aggregate (b) or remain attached (c) 
with probabilities $1-p_{1i}$ and $p_{1i}$, respectively.}
\end{center}
\end{figure}
\begin{equation}
{\partial \phi_{ij} \over \partial t} = D_{\phi} \bigtriangledown^2 \phi_{ij} - \rho \Theta_{ij} \phi_{ij} + \delta_{\phi}\phi_{0} - \delta_{\phi}\phi_{ij}.
\end{equation}

The Heaviside function $\Theta_{ij} $ is used to indicate the presence or absence of cells in that 
particular patch of the lattice (so we have $\Theta_{ij} =1$ if a cell is present and zero otherwise). 
On this same term, the parameter $\rho$ represents the intake rate of nutrients from the culture medium.
The last two terms of the equation are introduced as a replenishment process to ensure that, in the absence 
of cells, nutrient field recovers its initial value $\phi_{0}$. Here the diffusion operator 
$\nabla^2 \phi_{ij}$ is numerically computed (using the NetLogo libraries) by means of a standard discretization form: 

\begin{equation}
D_{\phi} \nabla^2 \phi_{ij} = D_{\phi} \left [\phi_{ij} - {1\over4} \sum_{kl} \phi_{kl}  \right ],
\end{equation}

where $D_{\phi}$ accounts for the diffusion coefficient. The energy change for i-th cell in the j-th aggregate is: 

\begin{equation}
{\partial M_{ij} \over \partial t} = \rho \phi_{ij} - \beta_c M_{ij}(1+\kappa \Delta_{ij}).
\end{equation}

Here $\beta_c$ represents the maintenance costs and $\Delta_{ij}$ accounts for the number of divisions 
this particular cell has undergone, causing cells to increasingly spread their divisions. If the energy value 
of a particular cell reaches its division threshold, a new cell is created and the original energy value is split 
asymmetrically between the cells. Conversely, cells also generate generic waste as a byproduct of their 
metabolic activity. The change in finite element $W_{ij}$ is:

\begin{equation}
{\partial W_{ij} \over \partial t} = D_{W} \bigtriangledown^2 W_{ij} + \gamma \Theta_{ij} M_{ij} - \delta_{W}W_{ij}.
\end{equation}

Similarly to the nutrient concentration, waste is created in those positions of the lattice occupied by cells 
(heaviside function $\Theta_{ij}$), in a quantity proportional to the maintenance costs of the cell ($\gamma M_{ij}$).
Waste is also subject to diffusion and decays proportionally to the current amount. 
Cells initiate apopotosis if the following threshold condition is met: 
$W_{ij} \geq \delta_c$, where $\delta_c$ is the upper bound that cells can withstand. 

In figure 3 we show an example of how aggregates grow in size, with increasing levels of waste until
some cells meet this threshold and die. As a result, a few smaller aggregates are created, which can 
export -through passive diffusion- enough waste to avoid death. This pattern of growth until a critical size
has been reached appeared quite robust to parameter changes (listed in the caption of figure 3), which 
were arbitrarily chosen and have arbitrary units.

\begin{figure*}[htbp]
\begin{center}
\includegraphics[width=0.65 \textwidth]{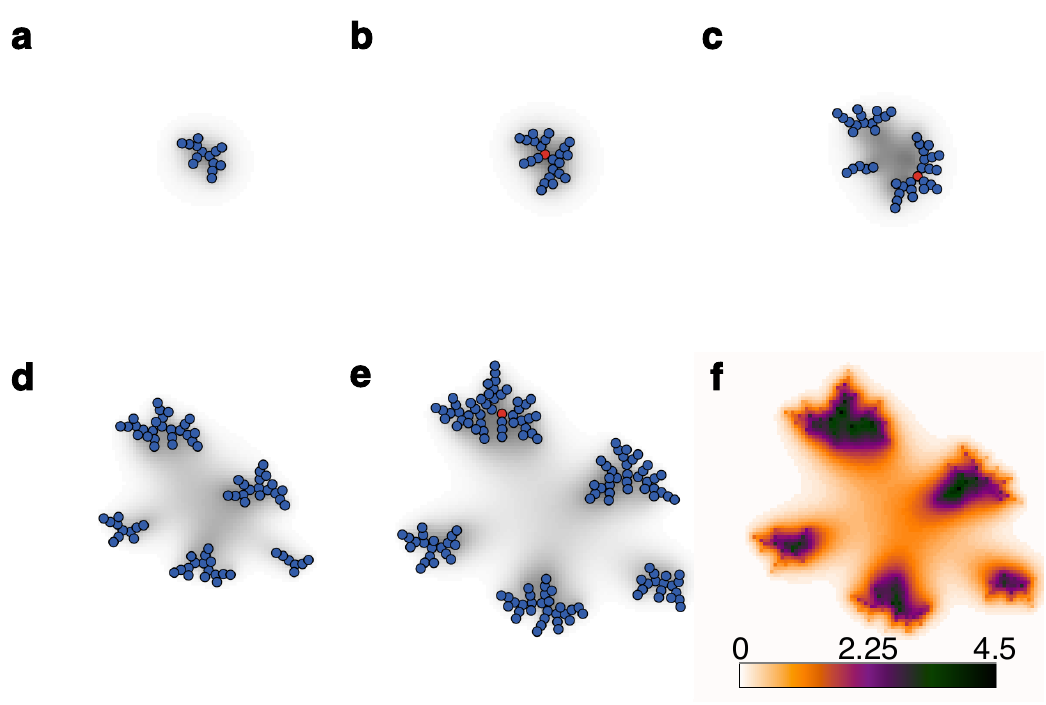}
\caption{Growth and sequential splitting of an aggregate due to cell death.  
Here (a-e), five spatial snapshots are shown at different times within the in silico growth phase.
Living and apoptotic cells are shown as blue and red circles respectively. After a fixed number of algorithm
cycles, apoptotic cells disappear with their springs, causing the breakage of the aggregate into smaller clusters (b,c and e).
The waste field appears as a  continuously shaded gradient, darker areas indicating higher 
concentration levels. As expected, the core of the aggregate is highly enriched in byproducts, 
eventually causing the death of cells when it surpasses a certain threshold. A different view of the 
waste field of snapshot e is given in (f). The simulation times in algorithm cycles for the snapshots 
are: 200, 250, 300, 350 and 400. The parameter's values used in this simulation are: $D_{\phi}=0.1$, 
$D_{W}=0.1$, $\rho=0.1$, $\beta_{c}=0.01$, $M_{c}=80$, $\kappa=0.25$, $\gamma = 0.2$, $\delta_{c}=4.5$, 
$\delta_{\phi}=0.1$, $\delta_{W}=0.01$, initial concentrations $\phi_0=20$ and $W_0=0$,  and a non-evolving 
adhesion probability $p=1$.}
\end{center}
\end{figure*}

\subsection{Mutation}
Little is known about the genetic changes behind the establishment of the snowflake phenotype 
reported in Ratcliff et al. Whether it involved extensive rewiring of basic adhesion toolkit genes or 
slight tuning of interactions in gene networks we do not know, but experiments involving different 
sedimentation times clearly show that correct separation between cells is not a binary, all or nothing, process.

In order to make the less assumptions about the genetic changes taking place in Ratcliff et al., 
our model enables evolution of only one cell parameter: $p_{ij}$, which stands for the probability 
of remaining attached to the offspring in the event of a division, and condenses the effect of 
multiple genes related to adhesion mutating independently. As such, $p_{ij}$ is a continuous 
variable constrained between zero and one. This parameter is inherited by daughter cells with 
very small variations, namely, a flat distribution $\pm$0.05 is applied at each division event.

\subsection{Selection process}
In Ratcliff's et al. paper, the researchers made use of gravity as the external force facilitating the 
differential deposition of cell aggregates [15]. Physically this corresponds to a 
simple property of increasingly large objects falling within a fluid medium 
with a given friction and a fixed gravity field.  In our model, we have used a simplified two step process
to emulate the experimental setup used by Ratcliff et al. 

At the beginning of each simulation a set of cells was created with random positions in the virtual 
space, and were grown until $500$ cells were obtained under agitation conditions (no sedimentation). 
Afterwards, we let the cells / aggregates fall until a fixed number simulation cycles had passed, the 
aforementioned settling time. Then, individues located at the bottom, below a given critical height $h_c$,  
were uplifted to new random positions leaving intact their history and traits. Moreover, the virtual 
medium was refreshed to homogeneous nutrient ($\phi_0$) and waste ($W_0$) level.

When growing, cells move in a random-walk 
fashion, which is an approximation to continuously shaken media. Since they have no preferential 
direction of movement they tend to be homogeneously distributed through the simulation space. When 
settling, we use a biased random-walk as an approximation of a sedimentation process. The bias introduced 
is computed using stroke's law. During this phase all cells/clusters 
tend to go towards the bottom, eventually reaching the selection zone. 
  
\begin{figure*}
\begin{center}
\includegraphics[width=0.75 \textwidth]{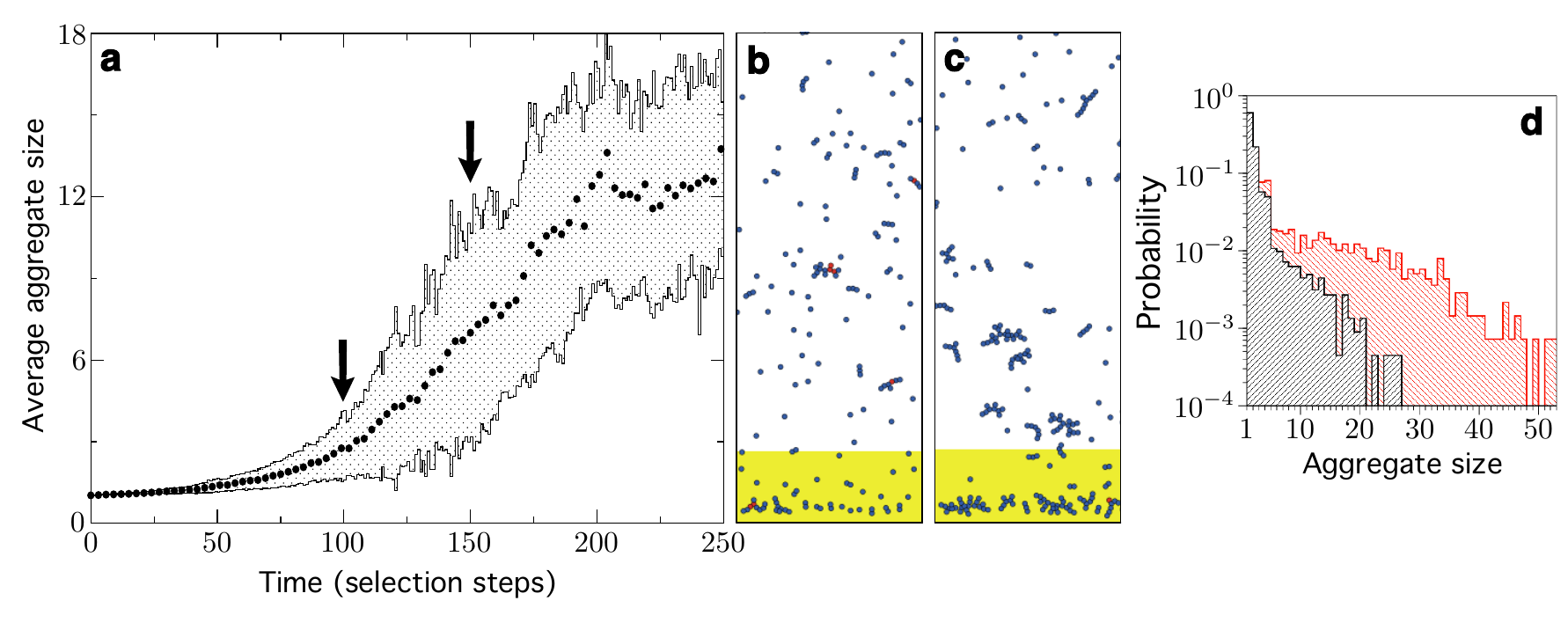}
\caption{(a) Time evolution of average aggregate sizes using the same parameters described 
before (figure 3) but with free, evolving $p_i$. Values shown (black dots) are averages of 30 replicate 
experiments and the shaded area represents one standard deviation of the dataset. In (b, c) 
we show two snapshots, obtained at the end of transfers $100$ and $150$ 
respectively (marked by arrows in a), which correspond to $\langle p \rangle=0.25$ and 
$\langle p \rangle=0.75$ (data not shown). In (d) we also show the size distribution for $t=150$, both in the selected and non-selected area 
(red and black respectively).}
\label{figseries1_model}
\end{center}
\end{figure*}
 
\section{Results}

Several traits of the multicellular aggregates emerging through the simulation can be measured
 with the experimental results discussed above. In our study we have followed both average 
 values of aggregate size over generations as well as those selected traits (such as cell-cell 
 adhesion) favouring the selection process towards larger aggregates. We can estimate the 
 probability of finding aggregates of a given size $\vert A_i \vert$, given by:
 
\begin{equation}
P(\vert A_i \vert; t) = { N(\vert A_i \vert ; t) \over\sum_{\mu=1}^{M} N(\vert A_{\mu} \vert ; t )}.
\end{equation}

 In figure 4a we display the evolution of the mean aggregate size as a function of time, calculated from: 
 
\begin{equation}
\langle \vert A \vert (t) \rangle  = {\sum_{\mu=1}^{M} \vert A_i \vert N(\vert A_{\mu} \vert ; t) \over
\sum_{\mu=1}^{M} N(\vert A_{\mu}  \vert ; t )}.
\end{equation}

We can appreciate a logistic-like growth pattern, thus exhibiting attrition after a given 
number of steps. The standard deviation is also displayed as a shaded envelope arround 
the mean. Two snapshots of the aggregate spatial distribution at the end of two selection 
phases are shown. These correspond to transfers $100$ and $150$, marked by arrows in (a). 
In these particular transfers, $\langle p \rangle$ had reached $0.25$ (b) and $0.75$ (c). A yellow region 
is included as a visual help to difference the selected region. In (d) we display the size 
distribution of aggregate sizes above (black) and below (red) the critical height $h_c$ for 
$T = 150$ and $\langle p \rangle=0.75$. It is possible to appreciate the progressive displacement
 towards higher aggregate sizes in the selected region (yellow) as a result of the sedimentation process. 

A specially relevant result seems to support our view. In Ratcliff's paper, it was shown that 
a highly nonlinear correlation exists between the size of the aggregate and the fraction of cells 
undergoing death within them (figure 5a, inset). In a nutshell, what is observed is that little death 
is found under a given aggregate size whereas it rapidly grows once we cross this threshold. 
However, the similar nonlinearity is obtained in our evolution model, as shown in figure 5a (main plot), where 
we display the statistics of cell death against the size of the aggregates. A nonlinear relationship 
is also found in our model, which is due to the nonlinearities associated to the thresholds of 
survival as well as the nonlinear relationships due to the geometric constraints imposed by our 
system. 

\begin{figure*}
\begin{center}
\includegraphics[width=0.6 \textwidth]{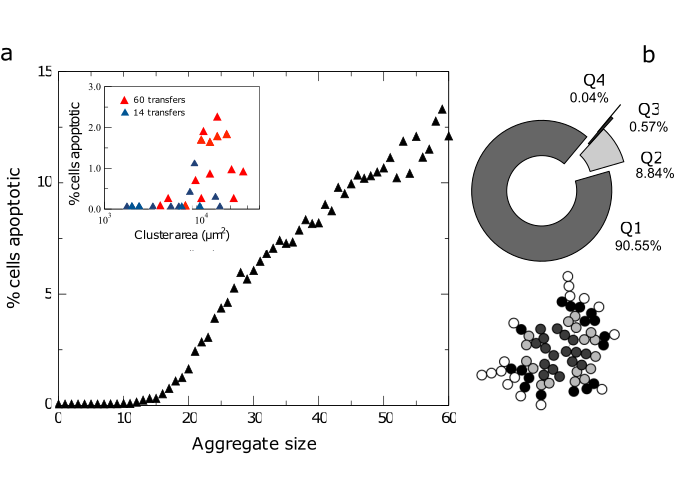}
\caption{Nonlinear relation between death rates and aggregate size (a). Our diffusion-driven 
model predicts a slow increase in death rates (here measured as the observed fraction of dead cells 
within a cluster) up to a certain aggregate size, from which death rapidly increases. Such a 
nonlinear relation was also found in the experiments (inset, adapted from Ratcliff et al 2012) and considered 
evidence for a selection process for programmed cell death. However (main plot) the diffusion-limited model 
discussed here predicts a very similar outcome, with low death rates below a given aggregate size 
and a sharp increase beyond that threshold. Localized cell death (b): $90.55\%$ of the $2\cdot10^4$ observed apoptotic cells  
were among the $25\%$ of cells closest to the centre of the aggregate. Below we show a sample cluster 
with its cells classified in four groups after their centrality, the different shades of grey 
correspond to the quartiles shown in the pie chart. }
\end{center}
\end{figure*}

\section{Discussion}
 
Unraveling the mechanisms responsible for the emergence of multicellular life forms 
out from single-cell systems represents a major challenge for our understanding of 
biological complexity.  The traditional approach to this problem was based either in 
data-driven, experimental and phylogenetic analysis or in mathematical and 
computer models of simple cell-like units and their emerging interactions [24]. 
The experimental work described in [15] provides a novel way 
of addressing this problem through a simple and elegant design of a selection-driven 
experimental setup. Despite the differences existing between wild and laboratory 
microorganisms [25] we can safely conjecture that the mechanisms responsible 
for generating and disaggregating cell clusters should be universal. 

Although the experimental results suggested an interpretation of the evolutionary 
dynamics in terms of an evolved, regulated response, the results reported here suggest a simpler, 
alternative interpretation in terms of a diffusion-limited process of aggregate 
growth where the cluster of cells keeps growing provided that enough waste 
is excreted passively into the medium. Once its size is large enough though, cells occupying the 
inner layers of the aggregate will start to trigger apoptotic mechanisms shown 
to occur in natural strains due to the accumulation of endogenous chemical cues, like acetic acid or
 ammonia. This alternative view does not disprove that the observed apoptotic rates
 are a consequence of adaptation, but offers a clear substrate from which evolution could act, 
 increasing the prevalence of an already existing mechanism.

Accordingly, our interpretation does not diminish the relevance and implications of the experimental 
evolution experiments. On the contrary, we think that this interpretation suggests 
a potentially interesting framework concerning the steps followed by primitive aggregates 
predating the first multicellular life forms. Aggregates breaking up due to internal cell 
death through toxicity results in a mechanism of splitting that clearly goes 
beyond the single-cell level, but is based in physical (or physico-chemical) constraints instead 
of actively operating regulatory mechanisms and signals. Such role played by physics over 
the cell's molecular machinery is consistent with a view of evolving multicellularity 
based on an early dominance of physical mechanisms over genetic ones [18,26-28]. 
 
Our model provides a simple computational framework that can be expanded in 
different ways. It also provides a useful system to design new forms of evolving 
multicellular aggregates. In this context, an interesting avenue can be considered here 
involving the use of synthetic biology, where specific engineered circuits 
for population size control or programmed cell death have been designed 
using microbial models. As a result of such work, it is fair to talk about to design cell-cell interactions in order to 
provide new, controlled scenarios of multicellular evolution [29]. In this context, we could 
take advantage of new engineered forms of cellular aggregation that can then be evolved 
over time. Such synthetic multicellular approach will offer a whole pathway of 
inquiry into the problem of how complex life might evolve or how we can evolve them.


\vspace{0.25 cm}

\noindent
{\bf Aknowledgments}

\vspace{0.25 cm}


We thank the members of the complex systems lab for useful discussions. RS thanks the European 
Center of Living Technology for its hospitality. This work has been supported by the
 Botin Foundation (SDN,RS), MINECO BES-2010-038940 fellowship 
(SDN) and by the Santa Fe institute, where most of the research was done. We would also like to 
thank the referees and editors for comments and suggestions that greatly improved the quality of the
manuscript.

\section{References}

\begin{enumerate}

\item
Bonner, J. T. 2001. First signals: the evolution of multicellular development. Princeton U. Press.

\item
Maynard Smith, J. and Szathm\'ary, E. 1995. The major transitions in evolution. 
Oxford U. Press.

\item
Erwin, D. H. 2008. Macroevolution of ecosystem engineering, niche construction and diversity.
Trends in ecology and Evolution. {\bf 6}, 304-310.

\item
Boraas, M.E.,  Seale D.B. and Boxhorn J.E. 1998. Phagotrophy by a flagellate selects for colonial prey: A possible origin of multicellularity. Evol. Ecol. Vol. {\bf 12}, 153-164.

\item
McShea, D.W. 2002. A complexity drain on cells in the evolution of multicellularity. Evolution {\bf 56}, 441-452. 

\item
King, N. 2004. The unicellular ancestry of animal development. Developmental Cell. {\bf 7}, 313-325.

\item
Grossberg, R.K. and and Strahmann, R.R. 2007. The Evolution of Multicellularity: 
A Minor Major Transition? Annu. Rev. Ecol. Evol. Syst. {\bf 38}, 621-654.

\item
Rokas, A. 2008. The molecular origins of multicellular transitions. Curr. Opin. Genet. and Dev. {\bf 18}, 472-478.

\item
Ruiz-Trillo, I., Roger, A. J., Burger, G., Gray, M. W., \& Lang, B. F. 2008. A phylogenomic investigation into the origin of metazoa. Molecular biology and evolution, 25(4), 664-672.

\item
Erwin, D. H. 2009. The early origin of the bilaterian developmental toolkit.
Phil. Trans. R. Soc. B. {\bf 364}, 2253-2261.

\item
Shapiro, J. A. 1998. Thinking about bacterial populations as multicellular organisms. Ann. Rev. Microbiol. {\bf 52}, 81-104. 

\item
Ben-Jacob E., Cohen I., Golding I., Gutnick D.L., Tcherpakov M., Helbing D. and Ron I.G. 2000. Bacterial cooperative organization under
antibiotic stress. Physica A. {\bf 282}, 247 - 282. 

\item
Seb\'e-Pedr\'os, A., Irimia M, del Campo J, Parra-Acero H, Russ C, 
Nusbaum C, Blencow BJ, and Ruiz-Trillo, I. 2013. Regulated aggregative multicellularity 
in a close unicellular relative of metazoa. eLife {\bf 2}, e01287.

\item
Olson, B. J. 2013. From brief encounters to lifelong unions. eLife {\bf 2}, e01893.

\item
Ratcliff, W.C., Denison, R.F., Borrello M. and Travisano M. 2012. 
Experimental evolution of multicellularity. Proc. Natl. Acad. Sci. USA. {\bf 109}, 1595-1600.

\item
Rebolleda-Gomez, M., Ratcliff, W.C. and Travisano, M. 2012.  Adaptation and Divergence during Experimental Evolution of Multicellular Saccharomyces cerevisiae. Artificial Life {\bf 13}, 99-104.

\item
Oud, B., Guagalupe-Medina, V., Nijkamp, J. F., de Ridder, D. et al. 2013. Genome duplication and mutations in ACE2 cause multicellular, fast-sedimenting phenotypes in evolved Saccharomyces cerevisiae. Proc. Natl. Acad. Sci. USA. {\bf 110}, E4223-E4231.

\item
Ratcliff, W.C., Pentz, J.T. and Travisano, M. 2013. Tempo and model of multicellular adaptation in experimentally evolved Saccharomyces cerevisiae. Evolution.{\bf 67}-6: 1573-1581

\item
Sol\'e, R.V.  and Valverde S. 2013. Before the endless forms: Embodied Model of Transition 
from Single Cells to Aggregates to Ecosystem Engineering. PLOS ONE {\bf 8}, e59664. 

\item
Ludovico, P., Sousa, M. J., Silva, M. T., Leao, C, Corte-Real, M. 2001. Saccharomyces cerevisiae commits to a programmed cell death process in response to acetic acid. Microbiology {\bf 147}, 2409-2415.

\item
V\'achov\'a, L. and Palkov\'a, Z. 2005. Physiological Regulation of Yeast Cell Death in Multicellular Colonies Is Triggered by Ammonia. The Journal of Cell Biology {\bf 169}, 711-717. 

\item
Carmona-Gutierrez, D., Eisenberg, T., Büttner, S., Meisinger, C., Kroemer, G., \& Madeo, F. 2010. Apoptosis in Yeast: Triggers, Pathways, Subroutines. Cell Death and Differentiation {\bf 17}, 763?73.

\item
Spears, W. M. and Spears, D. F.  2011. {\em Physico-mimetics: Physics-based swarm intelligence}. Springer, Heidelberg. 

\item
Kirk, D. L. 2005. A twelve-step program for evolving multicellularity and division of labor. Bioessays {\bf 27}, 299-310. 

\item
Palkov\'a A. 2004. Multicellular organisms: laboratory versus nature. EMBO reports {\bf 5}, 470-476.

\item
Newman, S. A., Forgacs, G. and M\"uller, G. B.  2006. Before programs: the physical origination of 
multicellular forms. Int. J. Dev. Biol. {\bf 50}, 289-299.

\item
Newman, S. A. and Baht, R. 2008. Dynamical patterning modules: physico-genetic determinants of 
morphological development and evolution. Phys. Biol. {\bf 5}, 015008.

\item
Sol\'e, R.V.  and Valverde S. 2013. Macroevolution in silico: scales constraints and universals. Paleontology {\bf 56}, 
1327Ð1340. 

\item
Maharbiz, M. M. 2012. Synthetic multicellularity. Trends Cell Biol. {\bf 12}, 617-623.

\end{enumerate}
\end{document}